\pdfoutput=1

\documentclass[11pt]{article}

\usepackage[preprint]{acl}

\usepackage{times}
\usepackage{latexsym}

\usepackage[T1]{fontenc}

\usepackage[utf8]{inputenc}

\usepackage{microtype}

\usepackage{inconsolata}

\usepackage{graphicx}
\DeclareRobustCommand{\senticon}[1]{\includegraphics[scale=0.02]{#1}}
\graphicspath{{figures/}}
\newcommand{\pnpl}{PNPL \includegraphics[height=1.5\fontcharht\font`\B]{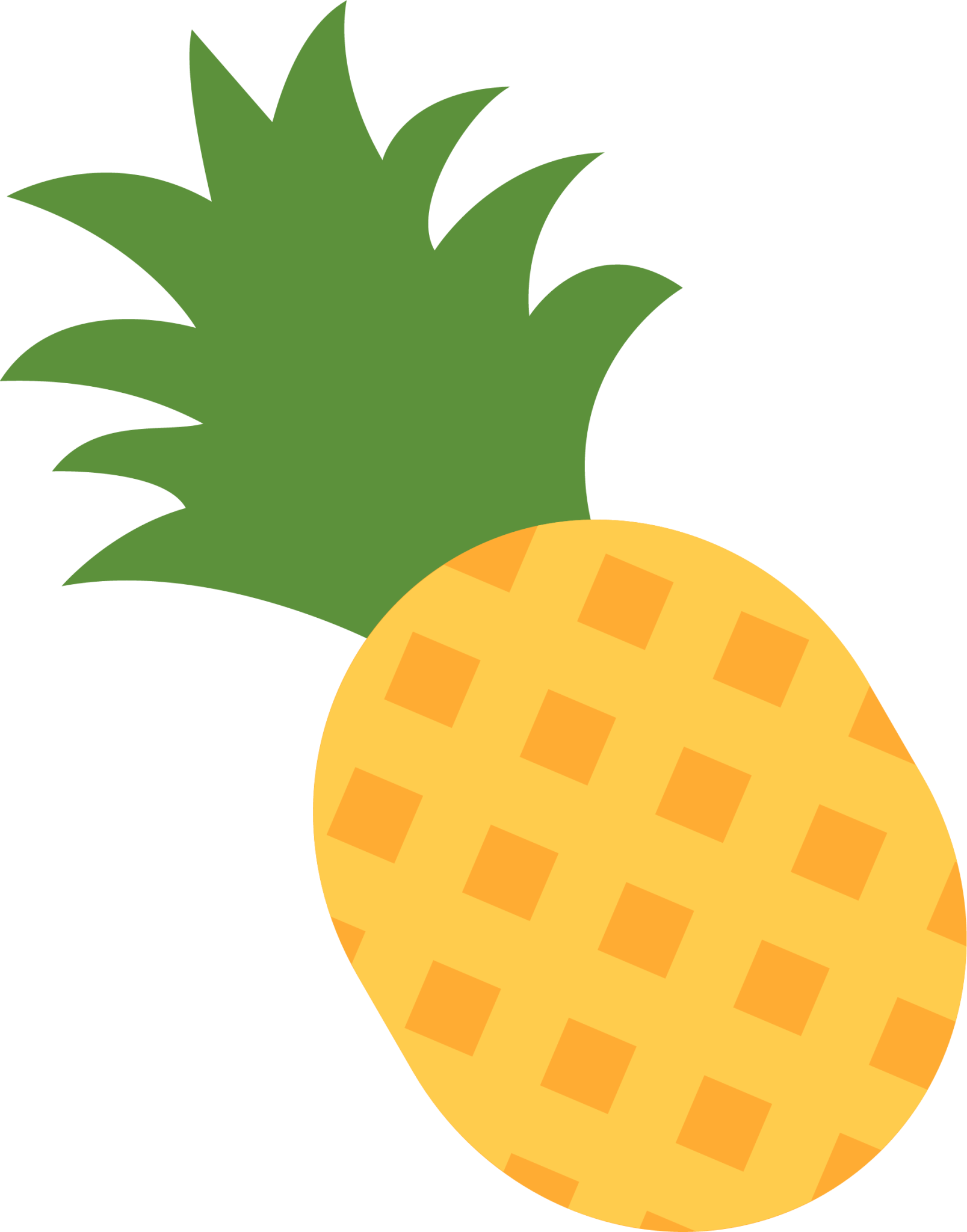}}

\usepackage{todonotes}

\usepackage{booktabs} %
\usepackage{stfloats} %

\title{MEGnifying Emotion: Sentiment Analysis from Annotated Brain Data} %

\author{Brian Liu \\
  \pnpl, University of Oxford, UK \\ %
  Bowdoin College, USA \\
  \texttt{bliu@bowdoin.edu} \\
  \And
  Oiwi Parker Jones \\
  \pnpl, University of Oxford, UK \\
  \texttt{oiwi@robots.ox.ac.uk} \\}

\begin{document}
\maketitle

\begin{abstract}
Decoding emotion from brain activity could unlock a deeper understanding of the human experience. While a number of existing datasets align brain data with speech and with speech transcripts, no datasets have annotated brain data with sentiment. To bridge this gap, we explore the use of pre-trained Text-to-Sentiment models to annotate non-invasive brain recordings, acquired using \emph{magnetoencephalography} (MEG), while participants listened to audiobooks. Having annotated the text, we employ force-alignment of the text and audio to align our sentiment labels with the brain recordings. It is straightforward then to train Brain-to-Sentiment models on these data. 
Experimental results show an improvement in balanced accuracy for Brain-to-Sentiment compared to baseline, supporting the proposed approach as a proof-of-concept for leveraging existing MEG datasets and learning to decode sentiment directly from the brain.
\end{abstract}

\section{Introduction}

\begin{figure*}[!b] %
  \centering
  \includegraphics[width=0.8\linewidth]{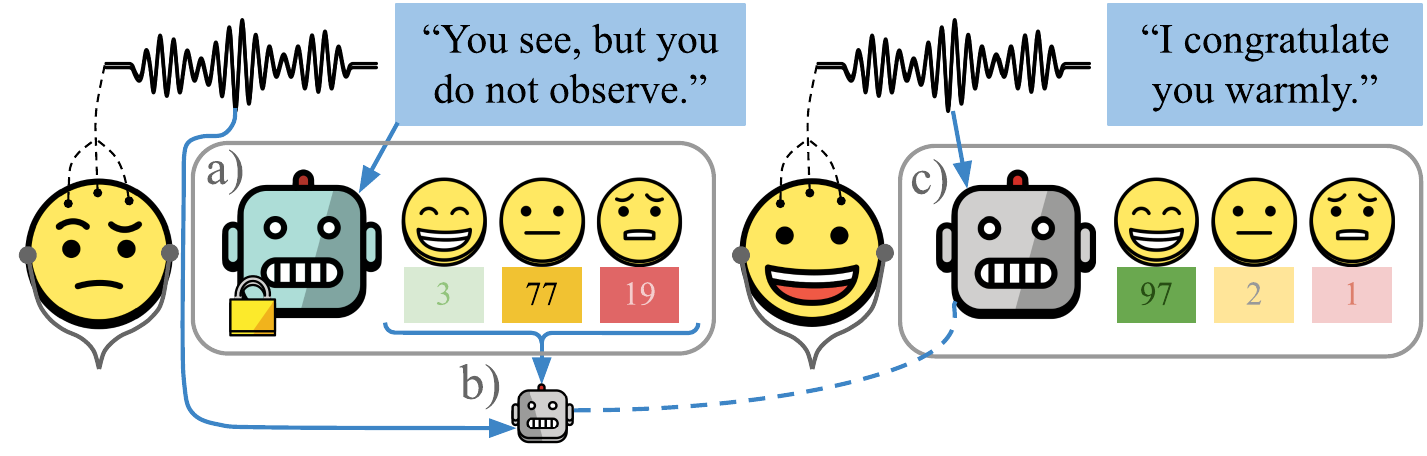}
  \caption{We utilize pretrained models to label text aligned with brain activity. Phrases of aligned text are given sentiment scores by a pretrained model (a), which are then used for training our model to predict sentiment from MEG (b). The model is evaluated by checking its predictions of unseen MEG data with the pretrained labels (c).}
  \label{fig:overview}
\end{figure*}

Neuroimaging techniques have seen dramatic improvements and increased documentation \cite{wintermark2018, GROSS2013349}, leading researchers to explore naturalistic paradigms to understand the brain \cite{hamilton2018revolution, NASTASE2020117254}. New brain imaging datasets that reflect this motivation are emerging, yielding an exciting time where deep learning techniques can directly decode information from the brain.

Decoding meaning from the brain has consequently been a focus of research in creating effective brain-computer interfaces. Recent MEG datasets include word and phoneme onset information, allowing for study of neural and textual relationships \cite{Armeni2022-hn, Gwilliams2023}. But there exist other contextual information informing brain activity such as emotional response. Yet, datasets with this information are few and far between. We propose leveraging pretrained Text-to-Sentiment models to label brain datasets with emotional response. With these labels, we hypothesize that it is possible to train a model to decode sentiment directly from MEG (Figure \ref{fig:overview}).

\section{Related works}

Differences in emotional processing between positive, neutral, and negative stimuli have been found by studying MEG channels, frequencies, and time ranges \cite{Kheirkhah2020-wx, Peyk2008-hw}, although predictive tasks have not been explored. Similarly, sentiment analysis from brain data has been widely explored \citep{DUTTA202325}, as well as directly decoding text from surgical and non-surgical methods \citep[e.g.~][]{Herff2015, moses_neuroprosthesis_2021, Metzger2023, Willett2023, Tang2023}.

Sentiment analysis with pretrained models from EEG data has been studied by \citealp{Wang_Ji_2022}, who used Text-to-Sentiment models to classify sentiment from decoded semantics. However, the results of this paper are contested as a flaw in the methodology may result in the method failing to beat a baseline that uses pure noise inputs \citep{jo2024eegtotext}.

\section{Sentiment labeling}

To the best of our knowledge, there are no MEG datasets that are labeled for sentiment. To create our own, we began with a relatively large MEG dataset collected from 3 subjects, each listening to 10 hours of an audiobook recording of \textit{The Adventures of Sherlock Holmes} \cite{Armeni2022-hn}. Our strategy takes the audio transcripts and labels the text using a pretrained sentiment analysis model.

The dataset includes event annotations (such as word onset times) obtained by text tokenization and forced alignment. However, there is no punctuation information to split the text for sentiment labeling. Using a transcript from other sources with punctuation information does not guarantee alignment with words in the event annotations because of minor differences in transcripts and normalization methods. For example, the event annotations in the dataset we use transcribes the audiobook recording of the year \verb|1852| as \verb|eighteen hundred fifty two| while other transcripts may read \verb|eighteen fifty two|. Cases such as these, in addition to the absence of punctuation, make aligning the transcripts non-trivial. Instead, we use natural pauses in the audiobook narration (marked as \verb|'sp'| in the event annotations) to divide the text into phrases rather than sentences.

\subsection{Sentiment analysis models}

There are many available off-the-shelf sentiment analysis models but they are trained on different datasets \textemdash we want to choose one that would generalize well to the text used in our MEG dataset. We started with a set of four publicly available models\footnote{Taken from https://huggingface.co/}, chosen based on popularity, target task (e.g. excluding any financial sentiment-analysis models), and our own initial tests. The models we compare (denoted by their author names or affiliations) are \verb|CardiffNLP| \citep{loureiro-etal-2022-timelms}, \verb|FiniteAutomata| \citep{pérez2023pysentimiento}, \verb|LXYuan| \citep{lik_xun_yuan_2023}, and \verb|NickWong| \citep{sheng-uthus-2020-investigating}. %
Each model outputs probabilities for a \verb|neutral|, \verb|positive|, and \verb|negative| class, solving a simple sentiment labeling problem.

\bgroup
\def\arraystretch{1.25}
\begin{table*}[t]
  \centering
  \begin{tabular}{l r r r c r c c}
    \toprule
    &\multicolumn{3}{c}{\textbf{Sentiment} (\%)}
    && \multicolumn{2}{c}{\textbf{Human correlation}} \\ 
    \cline{2-4} \cline{6-7}
    \textbf{Model} & 
    \multicolumn{1}{c}{\raisebox{-2.3px}{\includegraphics[scale=0.03]{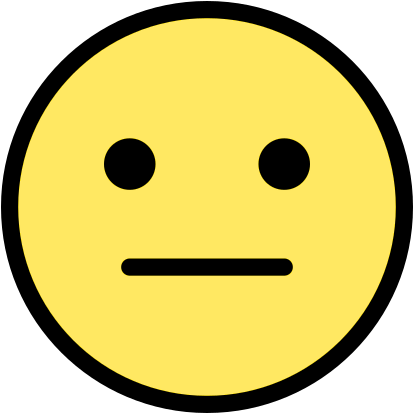}}} & 
    \multicolumn{1}{c}{\raisebox{-2.3px}{\includegraphics[scale=0.03]{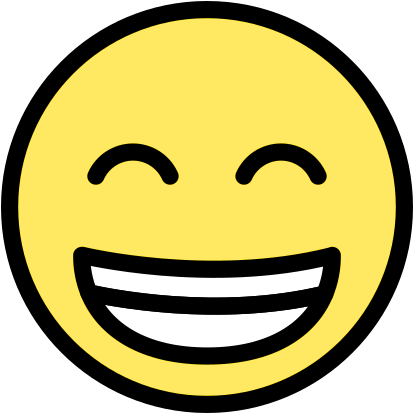}}} & 
    \multicolumn{1}{c}{\raisebox{-2.3px}{\includegraphics[scale=0.03]{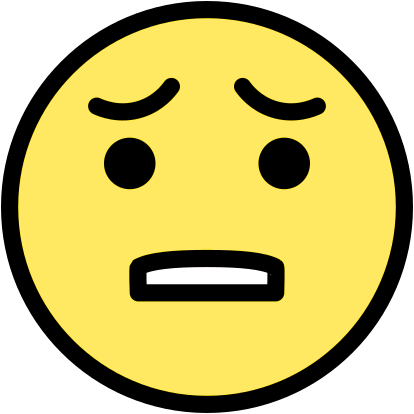}}} &&
    \multicolumn{1}{c}{$\rho$} &
    \multicolumn{1}{c}{$p$} &
    \multicolumn{1}{c}{\textbf{(Samples)}} \\
    \midrule
    {CardiffNLP}     & {85.048}        & {6.752}          & {8.200}           &&    {0.785}     & {0.0001} & {124M} \\
    {FiniteAutomata} & {85.983}        & {5.704}          & {8.313}           &&    0.701     & {0.0023} & {40K} \\
    {LXYuan}         & {3.462}         & {70.370}         & {26.168}          &&    0.672     & {0.0031} & {591K} \\
    {NickWong}       & {2.109}         & {57.331}         & {40.560}          &&    0.653     & {0.0045} & {296K} \\ 
    \bottomrule
  \end{tabular}
  \caption{Choosing a pre-trained sentiment analysis model to annotate brain recordings. We compare different pre-trained text-to-sentiment models on phrases from \textit{The Adventures of Sherlock Holmes}. \textbf{Sentiment (\%)} shows the proportion of phrases labeled neutral \protect\senticon{neutral.png}, positive \protect\senticon{positive.png}, and negative \protect\senticon{negative.png} for each model. \textbf{Human correlation} shows non-parametric correlations between model output and a human baseline, where $\rho$ is Spearman’s rank correlation coefficient and $p$ is the two-sided $p$-value. \textbf{Samples} shows the size of the dataset that each model is trained on.}
  \label{tab:models}
\end{table*}
\egroup

We use these models to infer the sentiment labels of the Sherlock text by phrases split by \verb|'sp'| and evaluate the proportion of each class. Our assumption is that the majority of the labels should be \verb|neutral| given that a realistic detective fiction primarily involves logical reasoning, detailed observation, and methodical deductions rather than emotional expressions or sentimental dialogue, which contrasts with other corpora (e.g. reviews). Table \ref{tab:models} reports the results for the four sentiment analysis models on our text corpus. We see that \verb|CardiffNLP|'s and \verb|FiniteAutomata|'s models best matched our expectations about the proportions of each class. 
But \verb|CardiffNLP| correlated better with a manually labeled subset of Sherlock phrases, as discussed in the next subsection.

\subsection{Correlation analysis between model predictions and human annotations}

We evaluated the correlation between sentiment labels produced by the four different pretrained models and human annotations from students and friends. The human annotations are provided as counts of \verb|neutral|, \verb|positive|, and \verb|negative| sentiments, while the model predictions are continuous values between 0 and 1. Each annotator was informed of the use of their data for comparison with pretrained models. To ensure a robust comparison, we use Spearman's rank correlation coefficient, which assesses the monotonic relationship between two variables.

We begin by aligning the data from two sources: the model outputs and the human annotations. 60 phrases were randomly sampled as a subset, and we obtained human annotations and the model outputs for each phrase. Spearman's correlation coefficient was then computed for each sentiment type (neutral, positive, negative), between a given model's output and the human annotated counts. The average correlation across the three sentiment types is used to determine the overall performance of each model.

The correlation results for each model are summarized in Table \ref{tab:models}. Among the four models, \verb|CardiffNLP|'s model exhibits the highest average Spearman correlation coefficient and a statistically significant $p$ value ($p \ll 0.05$, two-sided test), indicating the strongest alignment with human annotations. This suggests that \verb|CardiffNLP|'s model most effectively captures the sentiment expressed in the texts as perceived by human annotators.

As the \verb|CardiffNLP| both matches our expectations on the class proportions and correlates best with human judgements, in the rest of this paper, we use this model as our sentiment analysis model. These findings highlight the importance of selecting appropriate models for sentiment analysis tasks. The superior performance of \verb|CardiffNLP|'s model suggests that it may be more reliable for applications requiring high-fidelity sentiment detection.

\subsection{Data splits}

We split the sentiment-annotated MEG data into random 80\% train, 10\% validation, and 10\% test sets. Each split is associated with a random seed for repeated experimentation. 
\section{Predicting sentiment from the brain}

We now turn to the training and evaluation of models for predicting sentiment from MEG using the labels that we generated from the \verb|CardiffNLP| model and the existing audiobook transcripts.

\subsection{Evaluation metrics and baselines}

To evaluate the results, we use accuracy and balanced accuracy. For balanced accuracy, our baseline is the chance of guessing the correct class, which is 33.\={3}\% for three classes. For accuracy, the baseline would be the proportion of the majority class (\verb|neutral|), as always picking the majority class would be better than random guessing. 85.049\% of the training data belonged to the majority class.

The null hypothesis is that our best trained models will not be better than these metrics. The alternative hypothesis, which we measure using one-sided, one sample t-tests, is that the models will be better, at a 0.05 significance level. We use \texttt{scikit-learn} metrics to calculate our evaluation metrics.

\subsection{Model architectures and hyperparameters}

We explore our method with an MLP and RNN (LSTM) architecture. The output of \verb|CardiffNLP|'s model are 3 continuous values %
corresponding to the probability of the emotions. We learn to regress the probabilities output by this Text-to-Sentiment classifier. Using the probabilities rather than the sentiment classes as labels alleviates the difficulty of the classification task for the brain-to-sentiment classifier. This is particularly important given the low signal-to-noise ratio of brain data.

For each training loop, the input has shape $\mathrm{batch} \times \mathrm{time} \times \mathrm{channels}$, where $\mathrm{time}$ is sampled at 250 Hz and there were 269 sensor channels. For each architecture, we used the following set of hyperparameters:

\begin{itemize}
    \item MLP = 2 layers, 128 hidden units, 0.0001 Learning Rate, 32 batch size
    \item LSTM = 2 layers, 128 hidden units, 0.0001 Learning Rate, 32 batch size
\end{itemize}

For the MLP, we classify flattened $\mathrm{time} \times \mathrm{channels}$ tensors. This is known as full-epoch decoding in the neuroimaging literature \citep{csaky_interpretable_2023}. In contrast, the LSTM is a sequence model for which no flattening is required.

In each case, models are trained 10 times with unique random seeds for 200 epochs on the training data, evaluating loss on the validation set.

For our best architecture and hyperparameters, we evaluate the model on the held out test set against the baselines described above.

\section{Results}

\bgroup
\def\arraystretch{1.25}
\begin{table*}[tb]
  \centering
  \begin{tabular}{r c c c c}
    \toprule
    \textbf{Model} & \textbf{Accuracy} & \textbf{Balanced Accuracy} & \textbf{$t$} & \textbf{$p$} \\
    \midrule
    MLP          & $82.186 \pm 0.739$   & $35.878 \pm 0.335$ & $106.196$  & $1.475 \times 10^{-15}$  \\
    LSTM         & $87.371 \pm 0.201$   & $35.745 \pm 0.245$  & $
    144.588$  & \multicolumn{1}{r}{$9.2 \times 10^{-17}$}  \\
    \midrule
    Baseline & $85.049$ & $33.333$ & -- & -- \\
    \bottomrule
  \end{tabular}
  \caption{Brain-to-Sentiment results. Here we compare classification metrics across the baseline, MLP, and LSTM models. The MLP and LSTM results are averaged from 10 random seeds and show mean $\pm$ standard error. The statistics show one-sided one-sample t-tests.}
  \label{tab:results}
\end{table*}
\egroup

\begin{figure}[htb]
    \centering
    \includegraphics[width=\linewidth]{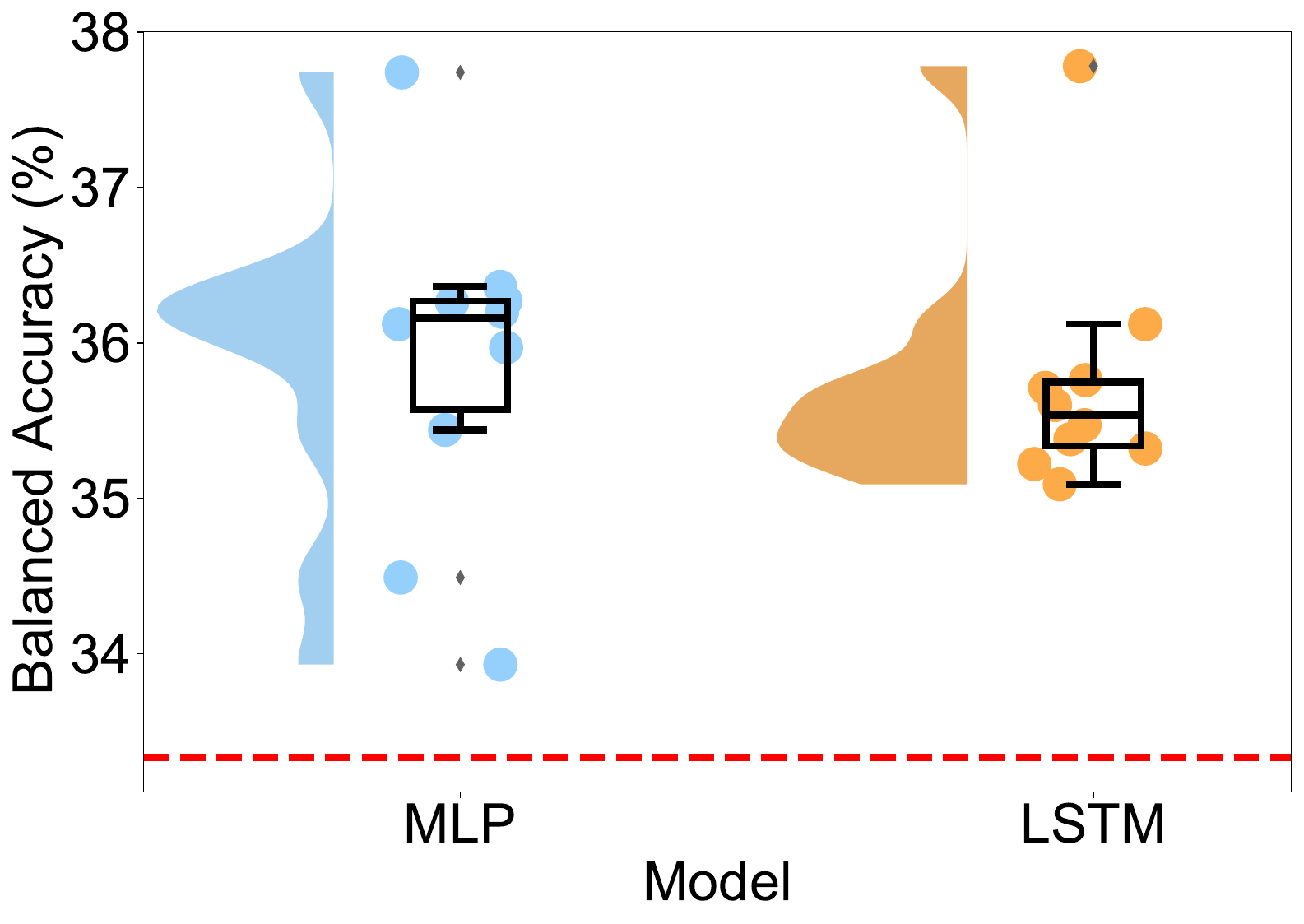}
    \caption{Distribution of balanced accuracy scores. The median balanced accuracy was larger for the MLP than for the LSTM predictions. However, the dispersion was also larger for the MLP than for the LSTM predictions. This explains why the statistics were more robust for the LSTM than for the MLP models. The red dashed line is the baseline which we would like the models to outperform.}
    \label{fig:rainclouds}
\end{figure}

On average, the largest accuracy came from the LSTM (87.371\%) rather than the MLP (82.186\%) (Table \ref{tab:results}). 
The mean balanced accuracy scores were much closer, with a slight advantage going to the MLP (35.878\% vs 35.745)\%). 
But the standard error was also larger for the MLP ($\pm 0.335$ vs $\pm 0.245$). 
Figure \ref{fig:rainclouds} illustrates the point, showing less dispersion of balanced accuracy scores for the LSTM. 

This explains why the best statistical results were obtained by the LSTM (Table \ref{tab:results}). 
Although both MLP and LSTM were better than guessing (33.333\%) when predicting positive, neutral, and negative sentiment labels ($p \ll 0.05$, one-sample t-tests), the effect size was larger for the LSTM ($t=144.588$) than for the MLP ($t=106.196$). 
We note, however, that we found no statistical difference in the MLP and LSTM results in a posthoc test ($t_{18}=0.321, p=0.752$, independent samples t-test).

\section{Discussion}

In this work, we explored leveraging pre-trained sentiment analysis models to annotate existing neuroimaging datasets. Having annotated a MEG dataset, we attempted to validate the method by training a Brain-to-Sentiment model. Annotating brain data with emotions can be difficult. How does one know what emotion to label and when? Using methods from NLP like sentiment analysis presents an objective and reproducible solution to labeling neural datasets that have already been aligned to speech.

Our experimental results demonstrate the viability of predicting sentiment directly from MEG. Although modest, the improvements in balanced accuracy over baseline performance support the proposition that deep learning can discriminate brain activity patterns related to sentiment (\verb|neutral|, \verb|positive|, \verb|negative|). 
The LSTM model showed the most consistent results both in terms of precision and effect size, perhaps indicating the potential of sequential models to capture temporal dependencies in MEG data.
But as a preliminary exploration into Brain-to-Sentiment, there are many things left to do about which we feel very positive.

\newpage
\section*{Limitations}
\label{sec:limitations}
Given the exploratory nature of this work, there are several limitations to consider. First, the difference in performance between the MLP and LSTM models was not substantial, suggesting that further hyperparameter tuning and model optimization could yield better results. There are many sequence models not explored in this work -- such as Transformers \citep{Vaswani2017Attention} and State Space Models \citep{Gu2022S4} -- that could perform better. 

Second, the high overall accuracy compared to balanced accuracy points to a class imbalance issue, where the model is biased towards predicting the majority class. 

This is highlighted by error analyses, where ground truth positive stimuli like \texttt{`and she could see that i was amused'} and ground truth negative stimuli like \texttt{`from some foolish freak when he was at college'} are predicted from brain data to be neutral. 
Future work could explore techniques such as oversampling, undersampling, or class-weighting to address this imbalance.

Third, we have not explored the interpretability of our results, in terms of the neural signals that underpin model performance. Positive, neutral, and negative affects are also vague. Future work could explore a richer set of sentiment labels, focusing for example on emotion-based sentiment analysis \citep{mohammad2010emotions, rasooli2021hugging, yang2022emobert, morabia2021emonet, bravo2016emoint}. 

Fourth, the dataset used for training and evaluation is relatively small, with only three subjects and a total of 30 hours of data. This limited sample size might not capture the full variability of brain responses to stimuli, which could impact the generalizability of our findings.

Fifth, the process of aligning text sentiment labels to MEG data relies on the accuracy of the pretrained sentiment analysis model and the precision of the event annotations. Any discrepancies in these alignments could introduce label noise into the training data, affecting model performance.

Sixth, the use of a single sentiment analysis model (CardiffNLP) might limit the diversity of sentiment representation. Exploring additional models or ensembling multiple models could provide a more nuanced understanding of sentiment in MEG data.

Lastly, the current models do not account for individual differences in brain activity. Personalized models that take into account subject-specific variations could potentially enhance prediction accuracy and provide deeper insights into the relationship between brain activity and sentiment.

Future research should address these limitations by incorporating larger and more diverse datasets, improving alignment techniques, and exploring personalized modeling approaches. Additionally, the integration of multimodal data, such as combining MEG with other neuroimaging modalities or physiological signals, could further enhance sentiment prediction capabilities.

\section*{Acknowledgments}

Many thanks to Dulhan Jayalath, Gilad Landau, as well as to other members of the PNPL group for constructive feedback. The authors would like to acknowledge the use of the University of Oxford Advanced Research Computing (ARC) facility in carrying out this work. \url{http://dx.doi.org/10.5281/zenodo.22558}. PNPL is supported by the MRC (MR/X00757X/1), Royal Society (RG\textbackslash R1\textbackslash 241267), NSF (2314493), NFRF (NFRFT-2022-00241), SSHRC (895-2023-1022), and ARIA (SCNI-SE01-P004).

\bibliography{anthology,custom}

@article{csaky_interpretable_2023,
	title = {Interpretable many-class decoding for {MEG}},
	volume = {282},
	issn = {10538119},
	url = {https://linkinghub.elsevier.com/retrieve/pii/S1053811923005475},
	doi = {10.1016/j.neuroimage.2023.120396},
	language = {en},
	urldate = {2023-11-02},
	journal = {NeuroImage},
	author = {Csaky, Richard and van Es, Mats W.J. and Jones, Oiwi Parker and Woolrich, Mark},
	month = nov,
	year = {2023},
	pages = {120396},
}

@article{jo2024eegtotext,
      title={Are {EEG}-to-Text Models Working?}, 
      author={Hyejeong Jo and Yiqian Yang and Juhyeok Han and Yiqun Duan and Hui Xiong and Won Hee Lee},
      year={2024},
      journal={arXiv},
      doi={https://arxiv.org/abs/2405.06459}
}

@misc{pérez2023pysentimiento,
      title={pysentimiento: A {Python} Toolkit for Opinion Mining and Social {NLP} tasks}, 
      author={Juan Manuel Pérez and Mariela Rajngewerc and Juan Carlos Giudici and Damián A. Furman and Franco Luque and Laura Alonso Alemany and María Vanina Martínez},
      year={2023},
      eprint={2106.09462},
      archivePrefix={arXiv},
      primaryClass={id='cs.CL' full_name='Computation and Language' is_active=True alt_name='cmp-lg' in_archive='cs' is_general=False description='Covers natural language processing. Roughly includes material in ACM Subject Class I.2.7. Note that work on artificial languages (programming languages, logics, formal systems) that does not explicitly address natural-language issues broadly construed (natural-language processing, computational linguistics, speech, text retrieval, etc.) is not appropriate for this area.'}
}

@misc {lik_xun_yuan_2023,
	author       = { {Lik Xun Yuan} },
	title        = { distilbert-base-multilingual-cased-sentiments-student (Revision 2e33845) },
	year         = 2023,
	url          = { https://huggingface.co/lxyuan/distilbert-base-multilingual-cased-sentiments-student },
	doi          = { 10.57967/hf/1422 },
	publisher    = { Hugging Face }
}

@article{GROSS2013349,
title = {Good practice for conducting and reporting {MEG} research},
journal = {NeuroImage},
volume = {65},
pages = {349-363},
year = {2013},
issn = {1053-8119},
doi = {https://doi.org/10.1016/j.neuroimage.2012.10.001},
url = {https://www.sciencedirect.com/science/article/pii/S1053811912009895},
author = {Joachim Gross and Sylvain Baillet and Gareth R. Barnes and Richard N. Henson and Arjan Hillebrand and Ole Jensen and Karim Jerbi and Vladimir Litvak and Burkhard Maess and Robert Oostenveld and Lauri Parkkonen and Jason R. Taylor and Virginie {van Wassenhove} and Michael Wibral and Jan-Mathijs Schoffelen}
}

@article{NASTASE2020117254,
title = {Keep it real: rethinking the primacy of experimental control in cognitive neuroscience},
journal = {NeuroImage},
volume = {222},
pages = {117254},
year = {2020},
issn = {1053-8119},
doi = {https://doi.org/10.1016/j.neuroimage.2020.117254},
url = {https://www.sciencedirect.com/science/article/pii/S1053811920307400},
author = {Samuel A. Nastase and Ariel Goldstein and Uri Hasson}
}

@incollection{DUTTA202325,
title = {Chapter 2 - A predictive method for emotional sentiment analysis by deep learning from EEG of brainwave dataset},
editor = {Ajith Abraham and Sujata Dash and Subhendu Kumar Pani and Laura García-Hernández},
booktitle = {Artificial Intelligence for Neurological Disorders},
publisher = {Academic Press},
pages = {25-48},
year = {2023},
isbn = {978-0-323-90277-9},
doi = {https://doi.org/10.1016/B978-0-323-90277-9.00002-X},
url = {https://www.sciencedirect.com/science/article/pii/B978032390277900002X},
author = {Pijush Dutta and Shobhandeb Paul and Korhan Cengiz and Rishabh Anand and Asok Kumar}
}

@article{Armeni2022-hn,
title     = "A 10-hour within-participant magnetoencephalography narrative
           dataset to test models of language comprehension",
author    = "Armeni, Kristijan and G{\"u}{\c c}l{\"u}, Umut and van Gerven,
           Marcel and Schoffelen, Jan-Mathijs",
journal   = "Sci. Data",
publisher = "Springer Science and Business Media LLC",
volume    =  9,
number    =  1,
pages     = "278",
month     =  jun,
year      =  2022,
copyright = "https://creativecommons.org/licenses/by/4.0",
url = {https://www.nature.com/articles/s41597-022-01382-7},
language  = "en"
}

@Article{Willett2023,
author={Willett, Francis R.
and Kunz, Erin M.
and Fan, Chaofei
and Avansino, Donald T.
and Wilson, Guy H.
and Choi, Eun Young
and Kamdar, Foram
and Glasser, Matthew F.
and Hochberg, Leigh R.
and Druckmann, Shaul
and Shenoy, Krishna V.
and Henderson, Jaimie M.},
title={A high-performance speech neuroprosthesis},
journal={Nature},
year={2023},
month={Aug},
day={01},
volume={620},
number={7976},
pages={1031-1036},
issn={1476-4687},
doi={10.1038/s41586-023-06377-x},
url={https://doi.org/10.1038/s41586-023-06377-x}
}

@Article{Tang2023,
author={Tang, Jerry
and LeBel, Amanda
and Jain, Shailee
and Huth, Alexander G.},
title={Semantic reconstruction of continuous language from non-invasive brain recordings},
journal={Nature Neuroscience},
year={2023},
month={May},
day={01},
volume={26},
number={5},
pages={858-866},
issn={1546-1726},
doi={10.1038/s41593-023-01304-9},
url={https://doi.org/10.1038/s41593-023-01304-9}
}

@article{wintermark2018,
    author = {Wintermark, Max and Colen, Rivka and Whitlow, Christopher T and Zaharchuk, Greg},
    title = "{The vast potential and bright future of neuroimaging}",
    journal = {British Journal of Radiology},
    volume = {91},
    number = {1087},
    pages = {20170505},
    year = {2018},
    month = {06},
    issn = {0007-1285},
    doi = {10.1259/bjr.20170505},
    url = {https://doi.org/10.1259/bjr.20170505},
    eprint = {https://academic.oup.com/bjr/article-pdf/91/1087/20170505/57380212/bjr.20170505.pdf},
}

@article{Herff2015, title={Brain-to-text: Decoding spoken phrases from phone representations in the brain}, volume={9}, DOI={10.3389/fnins.2015.00217}, journal={Frontiers in Neuroscience}, author={Herff, Christian and Heger, Dominic and de Pesters, Adriana and Telaar, Dominic and Brunner, Peter and Schalk, Gerwin and Schultz, Tanja}, year={2015}, month={Jun}}

@ARTICLE{Kheirkhah2020-wx,
  title    = "The Temporal and Spatial Dynamics of Cortical Emotion Processing
              in Different Brain Frequencies as Assessed Using the
              {Cluster-Based} Permutation Test: An {MEG} Study",
  author   = "Kheirkhah, Mina and Baumbach, Philipp and Leistritz, Lutz and
              Brodoehl, Stefan and G{\"o}tz, Theresa and Huonker, Ralph and
              Witte, Otto W and Klingner, Carsten M",
  journal  = "Brain Sci",
  volume   =  10,
  number   =  6,
  month    =  jun,
  year     =  2020,
  address  = "Switzerland",
  language = "en"
}

@ARTICLE{Peyk2008-hw,
  title     = "Emotion processing in the visual brain: A {MEG} analysis",
  author    = "Peyk, Peter and Schupp, Harald T and Elbert, Thomas and
               Jungh{\"o}fer, Markus",
  journal   = "Brain Topogr.",
  publisher = "Springer Science and Business Media LLC",
  volume    =  20,
  number    =  4,
  pages     = "205--215",
  month     =  jun,
  year      =  2008,
  language  = "en"
}

@article{Wang_Ji_2022, title={Open Vocabulary Electroencephalography-to-Text Decoding and Zero-Shot Sentiment Classification}, volume={36}, url={https://ojs.aaai.org/index.php/AAAI/article/view/20472}, DOI={10.1609/aaai.v36i5.20472}, abstractNote={State-of-the-art brain-to-text systems have achieved great success in decoding language directly from brain signals using neural networks. However, current approaches are limited to small closed vocabularies which are far from enough for natural communication. In addition, most of the high-performing approaches require data from invasive devices (e.g., ECoG). In this paper, we extend the problem to open vocabulary Electroencephalography(EEG)-To-Text Sequence-To-Sequence decoding and zero-shot sentence sentiment classification on natural reading tasks. We hypothesis that the human brain functions as a special text encoder and propose a novel framework leveraging pre-trained language models (e.g., BART). Our model achieves a 40.1% BLEU-1 score on EEG-To-Text decoding and a 55.6% F1 score on zero-shot EEG-based ternary sentiment classification, which significantly outperforms supervised baselines. Furthermore, we show that our proposed model can handle data from various subjects and sources, showing great potential for a high-performance open vocabulary brain-to-text system once sufficient data is available. The code is made publicly available for research purpose at https://github.com/MikeWangWZHL/EEG-To-Text.}, number={5}, journal={Proceedings of the AAAI Conference on Artificial Intelligence}, author={Wang, Zhenhailong and Ji, Heng}, year={2022}, month={Jun.}, pages={5350-5358} }

@Article{Gwilliams2023,
author={Gwilliams, Laura
and Flick, Graham
and Marantz, Alec
and Pylkk{\"a}nen, Liina
and Poeppel, David
and King, Jean-R{\'e}mi},
title={Introducing {MEG-MASC} a high-quality magnetoencephalography dataset for evaluating natural speech processing},
journal={Scientific Data},
year={2023},
month={Dec},
day={04},
volume={10},
number={1},
pages={862},
issn={2052-4463},
doi={10.1038/s41597-023-02752-5},
url={https://doi.org/10.1038/s41597-023-02752-5}
}

@Article{Metzger2023,
author={Metzger, Sean L.
and Littlejohn, Kaylo T.
and Silva, Alexander B.
and Moses, David A.
and Seaton, Margaret P.
and Wang, Ran
and Dougherty, Maximilian E.
and Liu, Jessie R.
and Wu, Peter
and Berger, Michael A.
and Zhuravleva, Inga
and Tu-Chan, Adelyn
and Ganguly, Karunesh
and Anumanchipalli, Gopala K.
and Chang, Edward F.},
title={A high-performance neuroprosthesis for speech decoding and avatar control},
journal={Nature},
year={2023},
month={Aug},
day={01},
volume={620},
number={7976},
pages={1037-1046},
issn={1476-4687},
doi={10.1038/s41586-023-06443-4},
url={https://doi.org/10.1038/s41586-023-06443-4}
}

@article{Vaswani2017Attention,
  title={Attention is All you Need},
  author={Vaswani, Ashish and Shazeer, Noam and Parmar, Niki and Uszkoreit, Jakob and Jones, Llion and Gomez, Aidan N and Kaiser, Łukasz and Polosukhin, Illia},
  journal={Advances in Neural Information Processing Systems},
  volume={30},
  year={2017}
}

@article{Gu2022S4,
  title={Efficiently Modeling Long Sequences with Structured State Spaces},
  author={Gu, Albert and Goel, Karan and R{\'e}, Christopher},
  journal={International Conference on Learning Representations (ICLR)},
  year={2022}
}

@inproceedings{mohammad2010emotions,
  author    = {Saif M. Mohammad and Peter D. Turney},
  title     = {Emotions Evoked by Common Words and Phrases: Using {Mechanical Turk} to Create an Emotion Lexicon},
  booktitle = {Proceedings of the NAACL HLT 2010 Workshop on Computational Approaches to Analysis and Generation of Emotion in Text},
  year      = {2010},
}

@inproceedings{rasooli2021hugging,
  author    = {Mohammad Sadegh Rasooli and Joel Tetreault},
  title     = {{Hugging Emotion Transformer}: A Pre-trained Model for Emotion Recognition in Text},
  booktitle = {Proceedings of the 2021 Conference on Empirical Methods in Natural Language Processing (EMNLP)},
  year      = {2021},
}

@inproceedings{yang2022emobert,
  author    = {Yiding Yang and Elizabeth Boschee and Tanya Goyal and Yoon Kim},
  title     = {{EmoBERT}: A Pretrained Language Model for Emotion Analysis in Text},
  booktitle = {Proceedings of the 2022 Conference of the North American Chapter of the Association for Computational Linguistics: Human Language Technologies (NAACL-HLT)},
  year      = {2022},
}

@article{morabia2021emonet,
  author    = {Keval Morabia and Francine Chen and Jeet Mohapatra},
  title     = {{EmoNet}: Large-scale Emotion Recognition for Multilingual {NLP}},
  journal   = {arXiv preprint arXiv:2105.13045},
  year      = {2021},
}

@inproceedings{bravo2016emoint,
  author    = {Felipe Bravo-Marquez and Mohammad Salameh},
  title     = {{EmoInt}: A Dataset and Benchmark for Emotion Intensity Recognition in Spanish},
  booktitle = {Proceedings of the 2016 Conference on Empirical Methods in Natural Language Processing (EMNLP)},
  year      = {2016},
}

@article{hamilton2018revolution,
  author    = {Hamilton, Liberty S. and Huth, Alexander G.},
  title     = {The revolution will not be controlled: Natural stimuli in speech neuroscience},
  journal   = {Lang Cogn Neurosci},
  volume    = {35},
  number    = {5},
  pages     = {573-582},
  year      = {2018},
  doi       = {10.1080/23273798.2018.1499946},
  PMID      = {32656294},
  PMCID     = {PMC7324135},
}

@article{moses_neuroprosthesis_2021,
	title = {Neuroprosthesis for {Decoding} {Speech} in a {Paralyzed} {Person} with {Anarthria}},
	volume = {385},
	issn = {0028-4793},
	url = {https://doi.org/10.1056/NEJMoa2027540},
	doi = {10.1056/NEJMoa2027540},
	number = {3},
	urldate = {2023-03-14},
	journal = {New England Journal of Medicine},
	author = {Moses, David A. and Metzger, Sean L. and Liu, Jessie R. and Anumanchipalli, Gopala K. and Makin, Joseph G. and Sun, Pengfei F. and Chartier, Josh and Dougherty, Maximilian E. and Liu, Patricia M. and Abrams, Gary M. and Tu-Chan, Adelyn and Ganguly, Karunesh and Chang, Edward F.},
	month = jul,
	year = {2021},
	pmid = {34260835},
	pages = {217--227},
}

\appendix
\clearpage

\end{document}